\documentclass[aps,prl,twocolumn,floats]{revtex4}
\usepackage{amsmath}
\usepackage{amssymb}

\usepackage{epsf,graphicx,color}

\newcommand{\eps}{\varepsilon}

\newcommand{\la}{\langle}
\newcommand{\ra}{\rangle}

\newcommand{\hC}{\hat{C}}

\begin{document}

\title{
Long-time signatures of short-time dynamics in decaying
quantum-chaotic systems}

\author{T. Gorin, D. F. Martinez}
\affiliation{Max-Planck-Institut f\" ur Physik komplexer Systeme,
        N\" othnitzer Str.\ 38, D-01187 Dresden, Germany}

\author{H. Schomerus}
\affiliation{Department of Physics, Lancaster University,
Lancaster, LA1 4YB, UK}

\date{\today}

\begin{abstract}
We analyze the decay of classically chaotic quantum systems in the presence 
of fast ballistic escape routes on the Ehrenfest time scale. For a continuous 
excitation process, the form factor of the decay cross section deviates from 
the universal random-matrix result on the Heisenberg time scale, i.e. for 
times much larger than the time for ballistic escape. We derive an exact 
analytical description and compare our results with numerical simulations for 
a dynamical model.
\end{abstract}

\maketitle

Properties of complex quantum systems are mostly characterized by a high
degree of universality which is rooted in the random interference of many
partial waves~\cite{GMW98}. A prototypical example are the cross section
fluctuations which arise from the random interference between the decay modes
of an excited quantum system. In the case of a large number of decay channels,
these fluctuations are free of all characteristics except for the classical
decay rate (Ericson fluctuations)~\cite{EriMay66,EngWei73,BluSmi88}.
The main theoretical tool to describe such universal properties is
random-matrix theory (RMT)~\cite{mehta,Bee97}. Semiclassical theories reveal 
limits to this description which are rooted in direct processes and hence best
characterized in the time domain. Most notably, if the complex quantum
dynamics is due to a chaotic classical limit, the random wave interference
is only established after a finite time, the so-called Ehrenfest
time~\cite{Ale96}. For shorter times, the quantum dynamics follows the
classical dynamics quasi-deterministically. Over the past years, a large number
of studies have explored the consequences of a finite Ehrenfest
time~\cite{Schom05,jacquod,brouwer,TiaLar04}, but almost exclusively these 
studies were concerned with stationary properties.

The purpose of this paper is to develop a dynamical theory for a physical
observable -- the form factor of the decay cross section of an excited quantum
system -- in which non-universal corrections can be detected directly in the
time domain. The dynamical theory is based on a stroboscopic description which
extends the previous analysis of stationary properties in systems with a
finite Ehrenfest time \cite{Jac03,Two03}. In the absence of direct decay
processes, our model exhibits the typical universal fluctuations found in
autonomous models of quantum decay and described by
RMT~\cite{AlhFyo98,FyoAlh98}. We incorporate quasi-deterministic direct decay 
by applying a mapping formalism which has earlier been used to describe Fano
resonances~\cite{Fano61} due to direct decay
processes~\cite{AFGIM03,GMI04,G05}. It turns out that the
deviations in the form factor from the universal behavior persist
up to times comparable to the Heisenberg time, hence far beyond
the regime of quasi-deterministic decay.

Molecular photo-dissociation and atomic auto-ionization are
examples of half collision processes (see Ref.\
\cite{halfcollision} and references therein), {\em i.e.} the
coherent dynamics of quantum systems which have been excited by
external means ({\em e.g.} the absorption of a photon) and decay
subsequently via energetically accessible open channels.
Within the half-collision description,
each decay amplitude is given by the overlap
$\langle\Psi_a|\alpha\rangle$ of the square integrable initial state
$|\alpha\rangle$ and the scattering state $|\Psi_a\rangle$ associated to
channel $a$. 
The form factor $\hat C(t)$ is defined as the Fourier transform of
the autocorrelation function of the total decay cross section
$\sigma(E)=\sum_a|\langle\Psi_a|\alpha\rangle|^2$. It can
also be obtained from the Fourier transform of the cross section
itself,
\begin{equation}
\hat C(t)= \frac{1}{L}\; |\hat \sigma(t)|^2,\quad \hat\sigma(t)=
\int_{-L/2}^{L/2}\!\!\!
   d E\; e^{-2\pi i\, Et}\; \sigma(E) \; .
\label{decayff2}\end{equation} 
Here, we measure time in units of the Heisenberg time
$t_H=h/\Delta$, and energy in units of the mean level spacing $\Delta$ ($h$ is
Planck's constant). Typically, only a finite energy
range, classically small but quantum mechanically large ($L\gg
1$), enters this formula.

In previous works
\cite{AlhFyo98,FyoAlh98,AFGIM03,GMI04,G05}, the random-matrix
description of half-collision processes has been set up
for autonomous systems which are described by an
effective Hamiltonian~\cite{GMW98}. In the case of time reversal invariance,
the autocorrelation function then follows from the
Verbaarschot-Weidenm{\"u}ller-Zirnbauer (VWZ) integral~\cite{VWZ85}, and for
systems without direct decay the form factor is given by
\begin{align}
C_0(t) &= \!\!\!\!\!\!\!\!\! \int\limits_{{\rm max}(0,t-1)}^t\!\!\!\!\!\!\!\!
   d r\; \int\limits_0^r\frac{d u}{2u+1}\;
   \frac{4\, (t-r)(r+1-t)^{1+N}\; f_0}
      {(t^2 - r^2 + x)^2 (1 + 2r + x)^{N/2}} \notag\\
x &= u^2\; \frac{2r+1}{2u+1}\qquad f_0= r^2+2rt+t-x \; .
\label{crmt}\end{align}

{\em Stroboscopic model.} We base our dynamical theory of
half-collision processes on a time-periodic quantum map
description which has been previously used to describe 
transport in systems with a finite Ehrenfest
time, $t_{\rm Ehr}$~\cite{Prange,Vallejos,FyoSom00,Jac03,Two03}.
Because of time-reversal invariance of the decay dynamics, the
Floquet operator $F$ is symmetric, $F=F^T$, which allows to
decompose $F$ into $F=F_{\rm out}\,F_{\rm in}$, with $[F_{\rm
in}]^T= F_{\rm out}$~\cite{mehta}. To introduce decay, we specify an
$N$-dimensional subspace (spanned by a column vectors of the
matrix $p$) within the $M$-dimensional Hilbert space of $F$. That
subspace provides the interface between the closed system and the
decay channels. For ballistic decay (ideal coupling)  we define
$Q= 1- pp^T$ and obtain the time-reversal symmetric open quantum
map \cite{Jac03,Two03}
\begin{equation}
\psi(n+1)=F_{\rm in} Q F_{\rm out}\; \psi(n)\; ,
\label{dissmap}\end{equation}
which describes how the internal
wave function $\psi$ is degraded sequentially in successive decay
attempts (where the integer $n$ is the stroboscopic time).

If we consider a short excitation pulse which leaves the system in the
internal initial state $|\alpha\ra = \psi(0)$, the quantum map~(\ref{dissmap})
yields
\begin{equation}
\psi(n)= \begin{cases}
       (F_{\rm out}^\dagger Q F_{\rm in}^\dagger)^{-n}\, |\alpha\ra
         &: n\le 0\\
       (F_{\rm in} Q F_{\rm out})^n\, |\alpha\ra &: n\ge 0
   \end{cases} \; ,
\label{hatsigma}\end{equation} for the forward and reversed time
evolution of the system. By contrast, for a continuous excitation
process, Eq.~(\ref{dissmap}) implies~\cite{unpub}
\begin{equation}
\sigma(\eps)= 1 + 2{\rm Re}\, \la\alpha|\,
   F_{\rm in} Q F_{\rm out}\,
   \frac{1}{e^{-i\eps} -  F_{\rm in} Q F_{\rm out}}\, |\alpha\ra \; .
\label{sigothm}\end{equation} Both types of experiments are
related to one-another by the fact that the Fourier transform of
$\sigma(\eps)$ is just the return amplitude. That is, in the energy and
time units of Eq.~(\ref{decayff2}):  
$\hat\sigma(t)= M \la\alpha|\, \psi(n)\ra$, where $t=n/M$. The stroboscopic 
form factor is then obtained from Eq.(\ref{decayff2}). Note that the spectrum
of $F$ is homogenous on the unit circle, so that we may set $L=M$.

{\em Direct processes.}
We now  incorporate quasi-deterministic direct processes by assuming that part
of the initial amplitude escapes
during the first iteration of the map. Note that the total cross
section contains the forward, but also the backward (time reversed)
evolution, Eq.~(\ref{hatsigma}). This means that the autocorrelation function
will be equally sensitive to direct decay
in either direction. To study this dependence quantitatively, we
assume the following decomposition for the initial state:
\begin{equation}
|\alpha\ra = |\alpha_0\ra + F_{\rm out}^\dagger\, p\, |\alpha^+\ra
+ F_{\rm in}\, p\, |\alpha^-\ra \; .
\label{adecomp}\end{equation} Here, $|\alpha_0\ra$ and
$|\alpha^\pm\ra$ are assumed to be real in the basis chosen. Also,
we assume that the three terms are approximately orthogonal to each other,
{\it e.g.} $\la\alpha^+|p^\dagger F p |\alpha^-\ra \approx 0$ which would
follow from the absence of any prompt processes in the full scattering 
system.%
\footnote{In order that $\la\alpha^+|p^\dagger F p |\alpha^-\ra$ is different
from zero, $F$ must map parts of the $p$-subspace onto itself. That implies
a prompt component in the full scattering process.}
Hence, we find:  $\|\alpha\|^2= \|\alpha_0\|^2 + \|\alpha^-\|^2 +
\|\alpha^+\|^2$.
While $|\alpha_0\ra$ leads to purely indirect decay,
$\|\alpha^\pm\|^2$ gives the probability for direct decay within
the first step of the open map, Eq.~(\ref{dissmap}),
forwards/backwards in time. If $|\alpha^+\ra = |\alpha^-\ra$,
$|\alpha\ra$ becomes real, and the whole decay process becomes
symmetric in time. The orthogonality of the
three terms in Eq.~(\ref{adecomp}) then implies that 
the maximum amount of direct decay (in forward
or backward evolution) is restricted to one half.

\begin{figure}[t]
\begin{center}
\includegraphics[width=\linewidth]{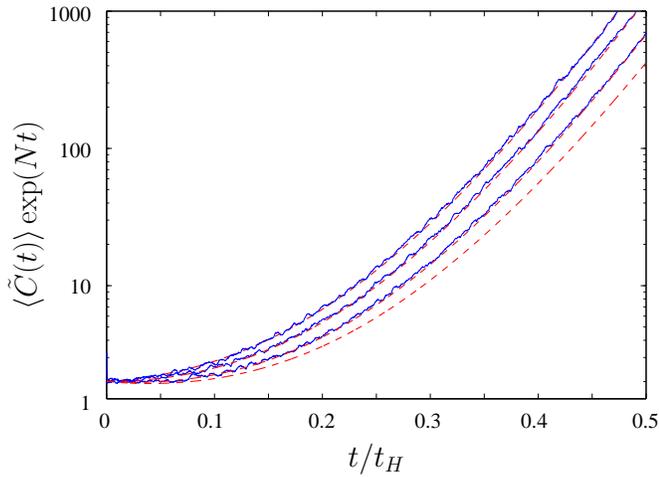}
\end{center}
\caption{\label{fig1}(color online)
Form factor $\la\tilde{C}(t)\ra$ times $\textrm{exp} (Nt)$ for the case of
time-reversal invariant decay (real initial coherent wave-packet in the open
kicked rotor) and for different amounts of  overlap of the initial state
with the regions of fast decay (solid lines).
Here, $M=4096$ and  $N=50$. The corresponding theoretical
results [Eq.(\ref{autoCt}), with $|\alpha^+\ra = |\alpha^-\ra$] are shown with 
dashed lines. From top to bottom $\|\alpha^\pm\|^2= 0, 0.25,0.417,0.5$.}
\end{figure}

To make contact with RMT we assume that $F_{\rm out}$ is 
taken from the circular unitary ensemble, which then implies that $F=
F_{\rm out} F_{\rm out}^T$ is a member of the circular orthogonal
ensemble~\cite{mehta}. The
resulting scattering ensemble \cite{FyoSom00}
is essentially equivalent to the random Hamiltonian ensemble discussed
in~\cite{GMW98}. In particular, the S-matrix correlations are well
described by the VWZ-integral~\cite{VWZ85,SFS05}. This allows to
calculate the autocorrelation function of the cross section in the
mapping formalism of Refs.~\cite{G05,GMI04}. The result is (for
details see \cite{tobepublished})
\begin{align}
&\hC (t)= \|\alpha_0\|^4\; C_0(t)\notag\\
&\quad + \frac{1}{2}\; \|\alpha_0\|^2 \big (
   \|\alpha^+\|^2 + \|\alpha^-\|^2 +
   2\, \la\alpha^+|\alpha^-\ra \big )\; C_{01}(t) \notag\\
&\quad
   + \frac{1}{4} \big ( \|\alpha^+\|^2\, \|\alpha^-\|^2
      + \la\alpha^+|\alpha^-\ra^2 \big )\; C_{11}(t) \; ,
\label{autoCt}\end{align}
where $C_{01}(t)$ and $C_{11}(t)$ are given by the expression~(\ref{crmt}),
but with the function $f_0$ replaced by $f_{01}= t$ and
$f_{11}= (x+r)/(1+2r+x) + (t-r)/(1+r-t)$, respectively. 
The auxiliary correlation functions $C_0(t)$, $C_{01}(t)$ and $C_{11}(t)$
all tend to two as $t\to 0$.
The functions $C_0(t)$ and $C_{11}(t)$ can be seen in Fig.~\ref{fig1}. They 
correspond to the limit cases of purely indirect ($\|\alpha^\pm \|^2=0$) and 
maximally direct ($\|\alpha^\pm\|^2=0.5$) decay. The function $C_{01}(t)$, if 
plotted, would lie in between. On a time scale given by $t_H\gg t_{\rm Ehr}$, 
all three functions are notably different.
Equation~(\ref{autoCt}) is the
central result of this paper. It generalizes
related results obtained within the autonomous random-matrix
model~\cite{GMI04,G05}. These results are recovered (for ballistic decay) by
choosing $|\alpha^+\ra = |\alpha^-\ra$, while the
universal random-matrix prediction~(\ref{crmt}) is obtained for 
$|\alpha^\pm\ra = 0$.
In general, the form factor depends on three independent parameters, the 
probabilities $\|\alpha^\pm\|^2$ for direct decay in forward/backward time 
evolution as well as the angle between the two corresponding amplitude vectors.

\begin{figure}[h]
\begin{center}
\includegraphics[width=\linewidth]{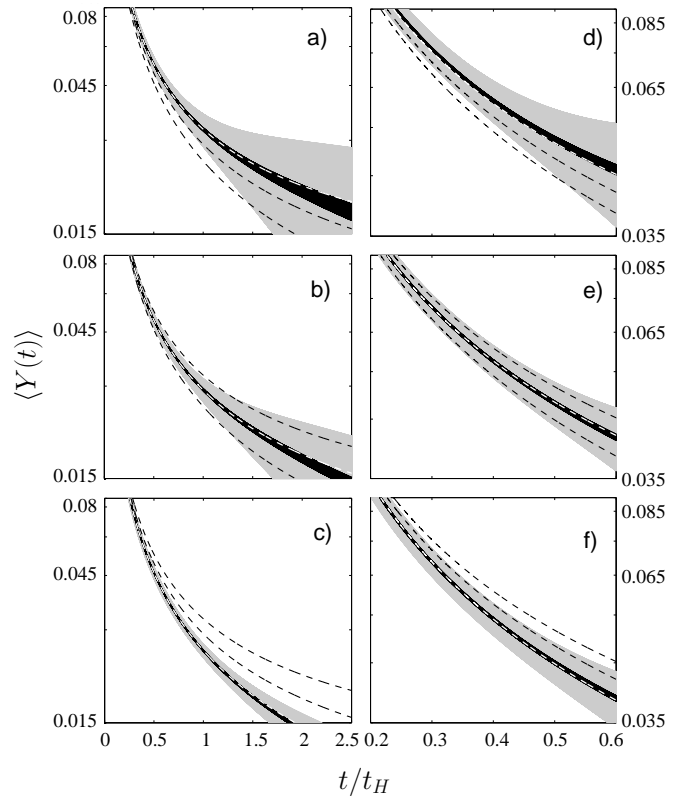}
\end{center} \caption{\label{fig2} Numerical data and
theory for $Y(t)$, for $N=20$ channels (left panels), and $N=50$
channels (right panels). The overlap of the initial state with the
regions of fast decay is given by $\|\alpha^-\|^2= 0$(a),
$0.595$(b), $0.907$(c), $0$(d), $0.571$(e), and $0.906$(f). 
Areas in gray represent the range of the standard deviation of $Y(t)$, while 
areas in black represent the standard deviation for $\left<Y(t)\right>$ when 
the average is taken over 100 samples.
The dashed lines show the theoretical curves for the different values for
$\|\alpha^-\|^2$; the respective theoretical curve corresponding
to the numerical result is plotted in light gray.}
\end{figure}

Maximally asymmetric decay is obtained by setting $|\alpha^-\ra$ to zero.
It yields:
\begin{equation}
\hC (t) = \|\alpha_0\|^4\; C_0(t) +
   \frac{1}{2}\; \|\alpha_0\|^2 \|\alpha^+\|^2\; C_{01}(t) \; .
\label{finalCt}\end{equation} This expression also describes the
decay form factor for an initial state $|\alpha\ra= F_{\rm in}\,
(|\alpha_0\ra + p\, |\alpha^-\ra)$, and yields the same dynamics
as choosing the initial state $|\alpha_0\ra + p\, |\alpha^-\ra$
for the open map $\chi(n+1)= Q\, F\; \chi(n)$ used {\it e.g.} in
Ref.~\cite{Two03,Schom04}.

{\em Numerical results.}
The decay form factor of realistic physical systems can be obtained from
experiments or in sophisticated numerical computations~\cite{StaWal05}. 
Here we test our theoretical predictions by numerical
investigations of a simple model system, the open kicked rotator
\cite{Jac03,Two03}, which was used in most studies of the
stationary properties of quantum systems with a finite Ehrenfest
time \cite{Schom05,jacquod,brouwer}. In the position basis the
Floquet matrix can be decomposed as $F= X\; U^\dagger\; \Pi\; U\;
X$, where
\begin{align}
\Pi &= {\rm diag}\big ( e^{-i\pi k^2/M}\big ), \notag\\
X &= {\rm diag}\big ( \exp[-i M\, V(2\pi k/M)\, ]\, \big ),
\end{align}
and $V(\theta)= K (\cos\theta -\gamma \sin\, 2\theta)$ with $\gamma =1$.
For $\gamma =0$, the standard kicked rotator, there are three different
symmetries present: time reversal, reflection and conjugation (the combination
of time reversal and reflection). A finite value of $\gamma$ breaks the 
reflection and also the conjugation symmetry. The initial wave-packet 
$|\alpha\ra$ is constructed as a superposition of two coherent states, located
around the line $p=0$ and therefore real. The amount of
overlap with the regions of fast decay (equal for both backward and forward
evolution) was controlled by changing the location of and/or squeezing the
coherent states.

We analyze the normalized autocorrelation function $\tilde C(t)=
2\; \hat C[\sigma](t)/ \hat C[\sigma](0)$, whose limit value at
small times is two, independent of the probabilities for direct
decay. This makes sure that the asserted sensitivity to direct
processes is related to the shape of the autocorrelation function,
not its norm (in practice, the cross sections measured or
calculated usually lack a reliable absolute scale).

We first consider the case of a real-valued initial state, with symmetric 
decay in time.  Figure~\ref{fig1} shows the  results for the form factor
$\la\tilde{C}(t)\ra$, scaled by the classical survival probability $\exp(-Nt)$. 
The numerical results are obtained for a large number of internal modes
$M= 4096$ (required for our random-matrix model to apply) and a large
number of open channels $N=50$. This allowed us to reach 
$\|\alpha^\pm\|^2= 0.417$ as the maximal probability for direct decay.
We take averages over 100 different realizations of the system, each
corresponding to a different value of the kick-strength $K$ (for all of them
the classical dynamics is completely chaotic).
The theoretical results derived from Eq.~(\ref{autoCt}) with parameters
corresponding to the numerical data are shown with dashed lines.
The numerical results agree perfectly well with the theoretical
predictions, within the remaining statistical uncertainty.

So far we have only discussed an ensemble average, while
experimentally or numerically it is not always possible  or
desirable  to change any system parameters. Can one detect the
direct quasi-deterministic decay for an individual system? This
question is addressed, for the case of maximally asymmetric decay, in 
Fig.\ \ref{fig2}. We show the results
for $M=10000$ and two values of $N$: $N=20$ in panels (a), (b), (c)
and  $N=50$ in panels (d), (e) and (f). On the vertical axis we
plot the function $Y(t)$ which we obtain from the numerical data
through the transformation~\cite{tobepublished}
\begin{equation}
Y(t)=\frac{1}{t}\sum_{m=20}^{Nt}
\frac{\tilde{C}(m/M)}{(C_0(m/M)-C_{01}(m/M))}\; ,
\label{Y}\end{equation}
which allows to obtain a best fit value for $\|\alpha^-\|^2$ by linear
regression (not shown).
In this sum we have discarded the first 20 kicks to exclude
system-specific processes at the Ehrenfest time scale.
It removes the fluctuations in time, so that only the
sample-to-sample fluctuations survive. These are shown by the gray
areas in Fig.~2, while the expected standard deviation for the
average over 100 samples is shown in
black. The dashed lines show the different theoretical curves, obtained
from Eq.~(\ref{finalCt}), that
fit best each one of the numerical results. Since the
sample-to-sample fluctuations (light gray) often do not cover the
whole available range between the theory for purely indirect and
dominantly direct decay, even a single experiment may be
sufficient to extract some information on the presence of direct
decay processes.

We have developed a dynamical description for the form factor $\hat C(t)$ of a 
decaying quantum system, taking into account the effect of direct 
(sub-Ehrenfest-time) decay processes. We derived a general analytical 
expression for $\hat C(t)$ at times of the order of the Heisenberg time, on 
the basis of a suitable random-matrix model.
While earlier studies~\cite{AlhFyo98,FyoAlh98,AFGIM03,GMI04,G05} implicitly
assumed that the wave packet will follow the same dynamics in forward and
backward evolution, we allow for excitation mechanisms which lead to 
asymmetric decay in time. In physical terms, asymmetric decay would result
from the preparation of an initial wave packet with finite group velocity. 
Properly directed this might strongly enhance the probability for prompt 
decay, but only in one direction in time.

We thank  A. Buchleitner and D. V. Savin for helpful discussions.
This work was supported by the European Commission, Marie Curie
Excellence Grant MEXT-CT-2005-023778 (Nanoelectrophotonics).

\end{document}